\documentclass[proceedings]{JHEP3}

\PrHEP{PrHEP hep2001}                   
\conference{International Europhysics Conference on HEP}                

\usepackage{epsfig}                   
\newcommand{\TeV}  {\ensuremath{\mathrm{TeV}}}
\newcommand{\GeV}  {\ensuremath{\mathrm{GeV}}}
\newcommand{\fb}   {\ensuremath{\mathrm{fb}}}
\newcommand{\invfb}{\ensuremath{\mathrm{fb^{-1}}}}
\newcommand{\PWm}  {\ensuremath{\mathrm{W^-}}}
\newcommand{\PWp}  {\ensuremath{\mathrm{W^+}}}

\newcommand{\PH}   {\ensuremath{\mathrm{H}}}
\newcommand{\PZ}   {\ensuremath{\mathrm{Z}}}
\newcommand{\Pep}  {\ensuremath{\mathrm{e^+}}}
\newcommand{\Pem}  {\ensuremath{\mathrm{e^-}}}
\newcommand{\epem} {\Pep\Pem}
\newcommand{\Pgne} {\ensuremath{\nu_{\mathrm{e}}}}
\newcommand{\Pagne}{\ensuremath{\overline{\nu}_{\mathrm{e}}}}
\newcommand{\Pq}{\ensuremath{\mathrm{q}}}
\newcommand{\Paq}{\ensuremath{\mathrm{\overline{q}}}}

\title{Electroweak Precision Tests with a Future Linear Collider}

\author{\speaker{Wolfgang Menges}
  Institut f\"ur Experimentalphysik, Universit\"at Hamburg, 
  Luruper Chaussee 149, 22761 Hamburg, Germany \\
        E-mail: \email{Wolfgang.Menges@desy.de}}           

\abstract{Future high energy linear electron positron colliders
  with centre-of-mass enegies between 90 and $1000\,\GeV$
  offer a unique opportunity to study precisely the Standard
  Model and phenomena from new physics. Most important, the 
  mechanism of electroweak symmetry breaking can be established
  in full depth if one or more Higgs bosons are observed.
  Precision measurements of electroweak processes open a window 
  into the multi-TeV energy range for searching for new physics,
  which is of particular importance if no light Higgs boson is 
  observed. }

\begin{document}

  \section{Framework}

  The physical motivation and the technical feasibility of linear
  electron-positron colliders (LC) has been intensively studied in 
  the past years in Japan (JLC \cite{Abe:2001gc}), the US 
  (NLC \cite{Abe:2001wn}) and Europe (TESLA \cite{Aguilar-Saavedra:2001rg}).
  The energy of such a machine would be $500\,\GeV$ in a first phase
  with upgrade potential to about $1\,\TeV$.

  A LC has a very rich physics potential covering both the study
  of new phenomena (Higgs physics, Supersymmetry and alternative physics) 
  and precision tests of the Standard Model (SM) (top quark physics, 
  electroweak physics and QCD).
  Here only prospects for precision Higgs measurements and
  precision electroweak measurements in the context of
  strong electroweak symmetry breaking (EWSB) will be described.
  The results presented here are based on integrated luminosities of
  500 to $1000\,\invfb$ and on realistic simulations of detector
  performance and background conditions.

  \section{Higgs Physics}

  In the SM and many of its extensions, EWSB proceeds via the
  Higgs mechanism leading to one or more physical Higgs bosons.
  The mass of the lightest Higgs boson is bounded from above on
  theoretical grounds, requiring perturbativity of the Higgs 
  coupling up to a high energy scale. Experimentally, 
  $m_{\mathrm{H}}<200\,\GeV$ is obtained from the interpretation 
  of electroweak precision measurements within the SM.

  While a Higgs boson, if it exists, is very likely to be found at
  Tevatron or LHC, the precise measurement of its properties and
  couplings can only be performed at the LC.

  The main production mechanisms for the SM-Higgs boson at the LC
  are Higgsstrahlung ($\epem\to\PZ\PH$) and WW-fusion 
  ($\epem\to\Pgne\Pagne\PH$). This gives $\mathcal{O}(10^5)$
  Higgs bosons per year for a light Higgs.
  Furthermore the Yukawa process $\epem\to\mathrm{t\bar{t}}\PH$ 
  and the double Higgsstrahlung process $\epem\to\PZ\PH\PH$ 
  can be exploited.
  The Higgs boson itself can be detected independent of its decay
  products through the observation of a mass peak in the invariant
  mass spectrum recoiling against a lepton pair from the $\PZ$-decay
  in the Higgsstrahlung process. Furthermore, all relevant Higgs
  boson decays ($\mathrm{b\bar{b}}$, $\mathrm{c\bar{c}}$, $\tau^+\tau^-$, 
  $\mathrm{gg}$, \PWp\PWm, \PZ\PZ, $\gamma\gamma$, $\mathrm{t\bar{t}}$)
  can be observed exclusively in the detector under study. 

  This large variety of accessible production mechanisms and decay
  modes in conjunction with the well defined initial state and clean 
  environment of \epem-collisions allows the determination of the 
  complete profile of the observed Higgs particle.

  The coupling $\mathrm{g_{HZZ}}$ of the Higgs boson to the \PZ{}
  can be derived in a model independent way from the observed 
  cross-section of the Higgsstrahlung process with leptonically 
  decaying \PZ-boson \cite{HiggsZH} (Fig. \ref{fig:hmm}). 
  The coupling $\mathrm{g_{HWW}}$
  can be derived from the cross-section for WW-fusion using 
  $\PH\to\mathrm{b\bar{b}}$ decays \cite{HiggsvvH}. This 
  cross-section can be disentangled from the Higgsstrahlung 
  contribution with $\PZ\to\nu\bar{\nu}$ exploiting the different 
  distributions of the missing invariant mass (Fig. \ref{fig:vvh}).

  \DOUBLEFIGURE[b]
  {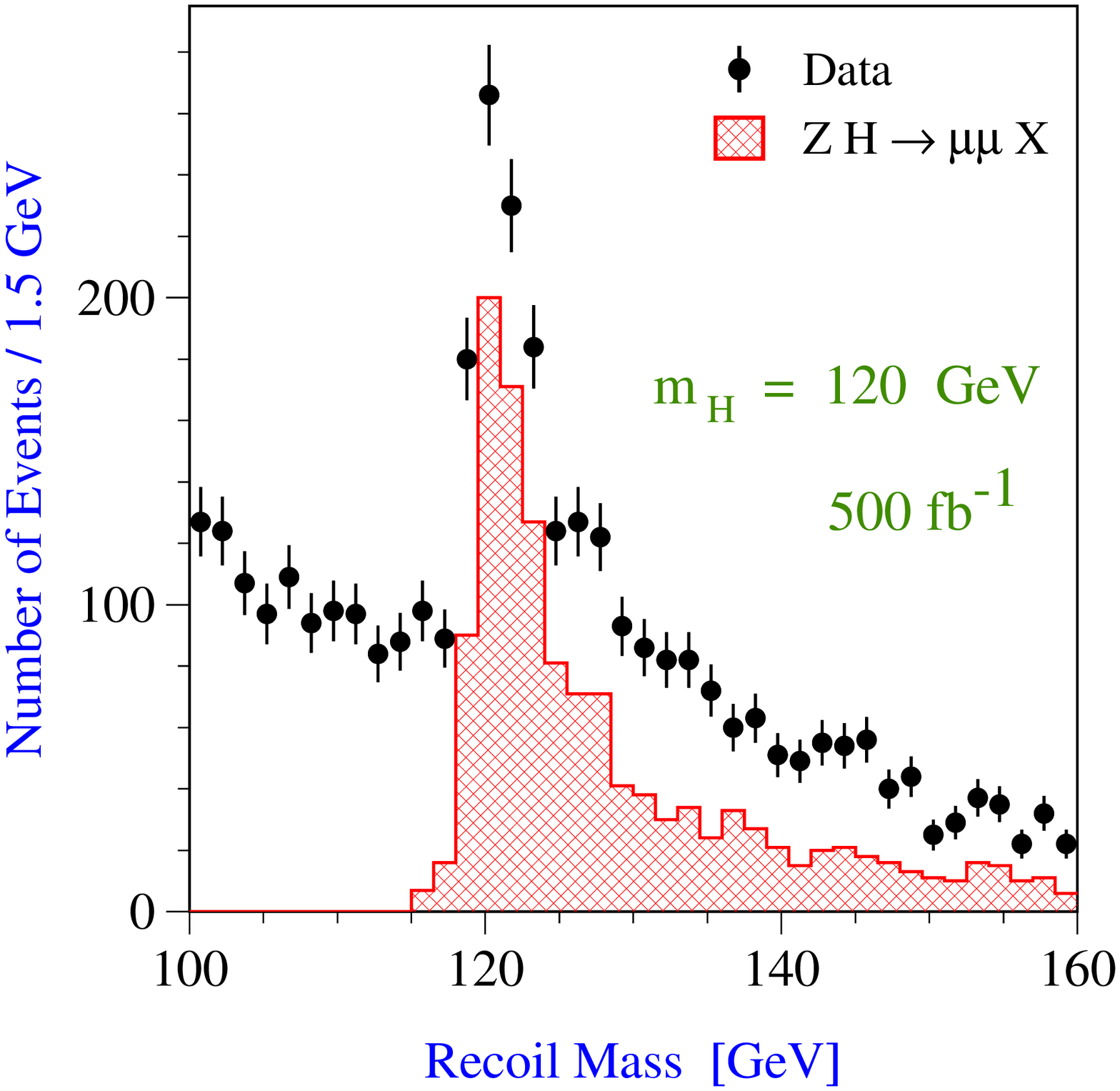,width=0.48\textwidth}
  {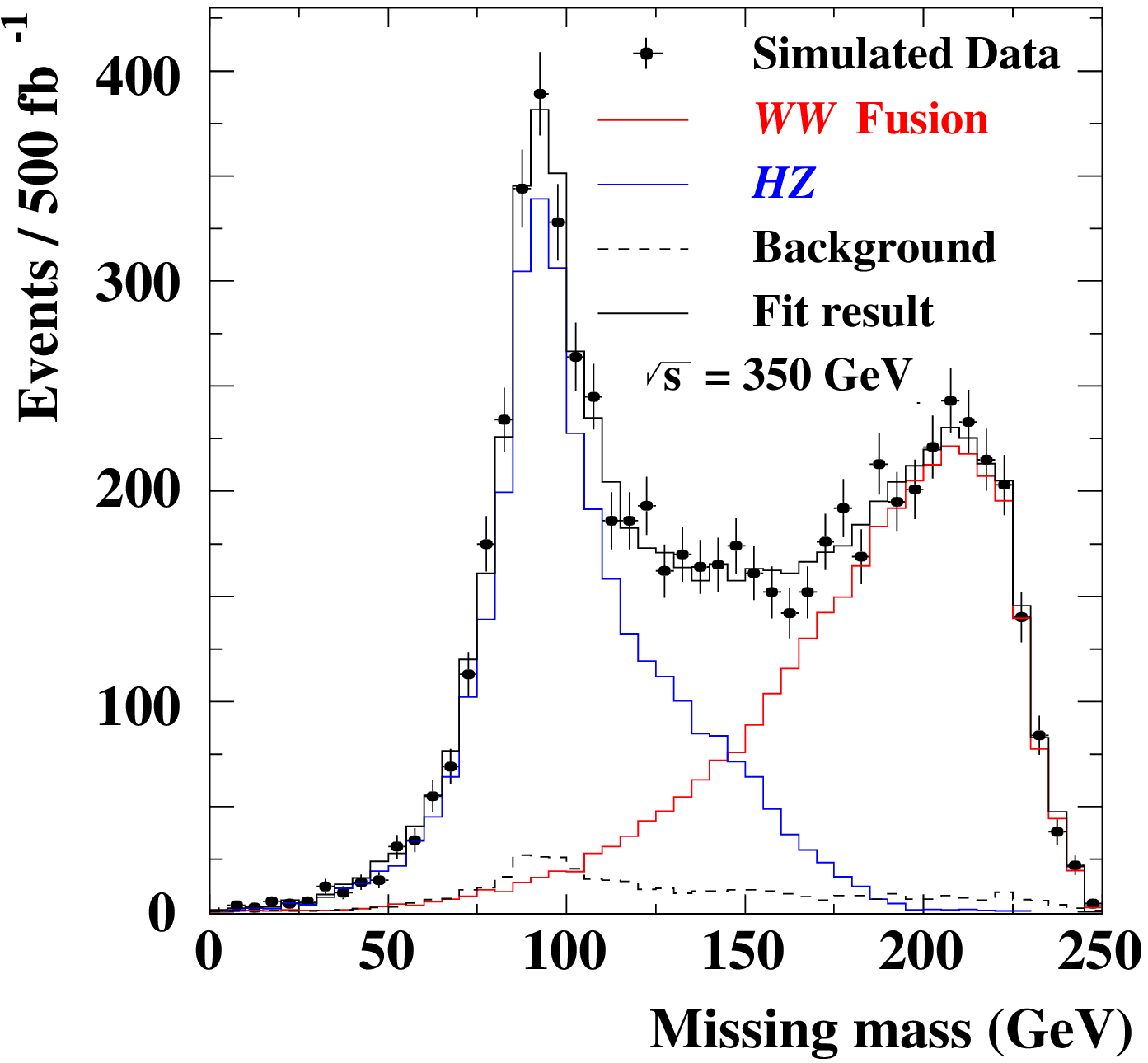,width=0.48\textwidth}
  {Higgs boson signal in the recoil mass distribution to a muon pair
    compatible with a \PZ{} decay for a simulated Higgs boson mass 
    $m_H = 120\,\GeV$ and $500\,\invfb$ at $\sqrt{s}=350\,\GeV$.
    \label{fig:hmm}}
  {Missing mass distribution in events with two b-jets with an
    invariant mass compatible with the Higgs boson mass for 
    $500\,\invfb$ at $\sqrt{s}=350\,\GeV$.
    \label{fig:vvh}}

  \EPSFIGURE
  {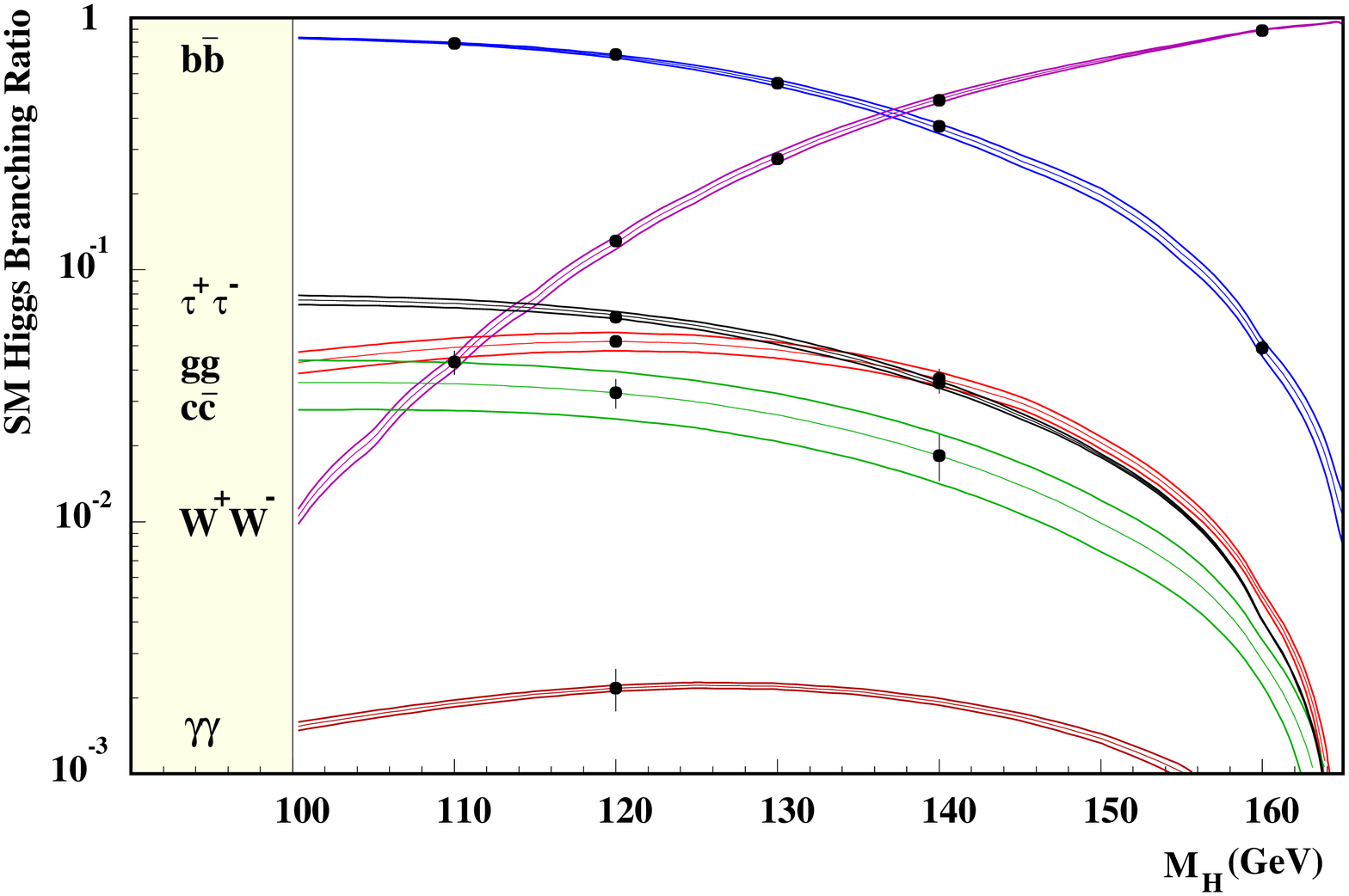,width=.6\textwidth}
  {The SM prediction for the decay branching ratios
    of the Higgs boson. The points with error bars
    indicate the achievable experimental precision
    for $500\,\invfb$. The bands indicate the theoretical
    uncertainties of the prediction.
    \label{fig:brhiggs}}

  An essential prediction of the Higgs mechanism is the Yukawa coupling
  of the Higgs boson to fermions being proportional to the fermion mass.
  This prediction can be accurately tested by measuring the Higgs boson
  decay branching ratios. The different hadronic Higgs boson decays
  $\mathrm{b\bar{b}}$, $\mathrm{c\bar{c}}$ and gg as well as
  $\tau^+\tau^-$ can be disentangled using the excellent flavour
  tagging capabilities of a LC detector \cite{HiggsBr}. 
  The expected precision of the branching ratio measurement is 
  ranging from $2\,\%$ (for $\mathrm{b\bar{b}}$) to approximately 
  $10\,\%$ (for gg) for a light Higgs boson ($m_{\mathrm{H}} < 160\,\GeV$) 
  (Fig. \ref{fig:brhiggs}). 

  The double Higgsstrahlung process $\epem\to\PZ\PH\PH$ gives
  access to the triple Higgs coupling $\lambda_{\mathrm{HHH}}$
  of the Higgs boson.   This coupling determines the shape of the 
  Higgs potential which is exactly predicted in terms of the 
  Higgs mass in the SM and thus provides a rigorous test of the 
  Higgs mechanism. 
  \EPSFIGURE
  {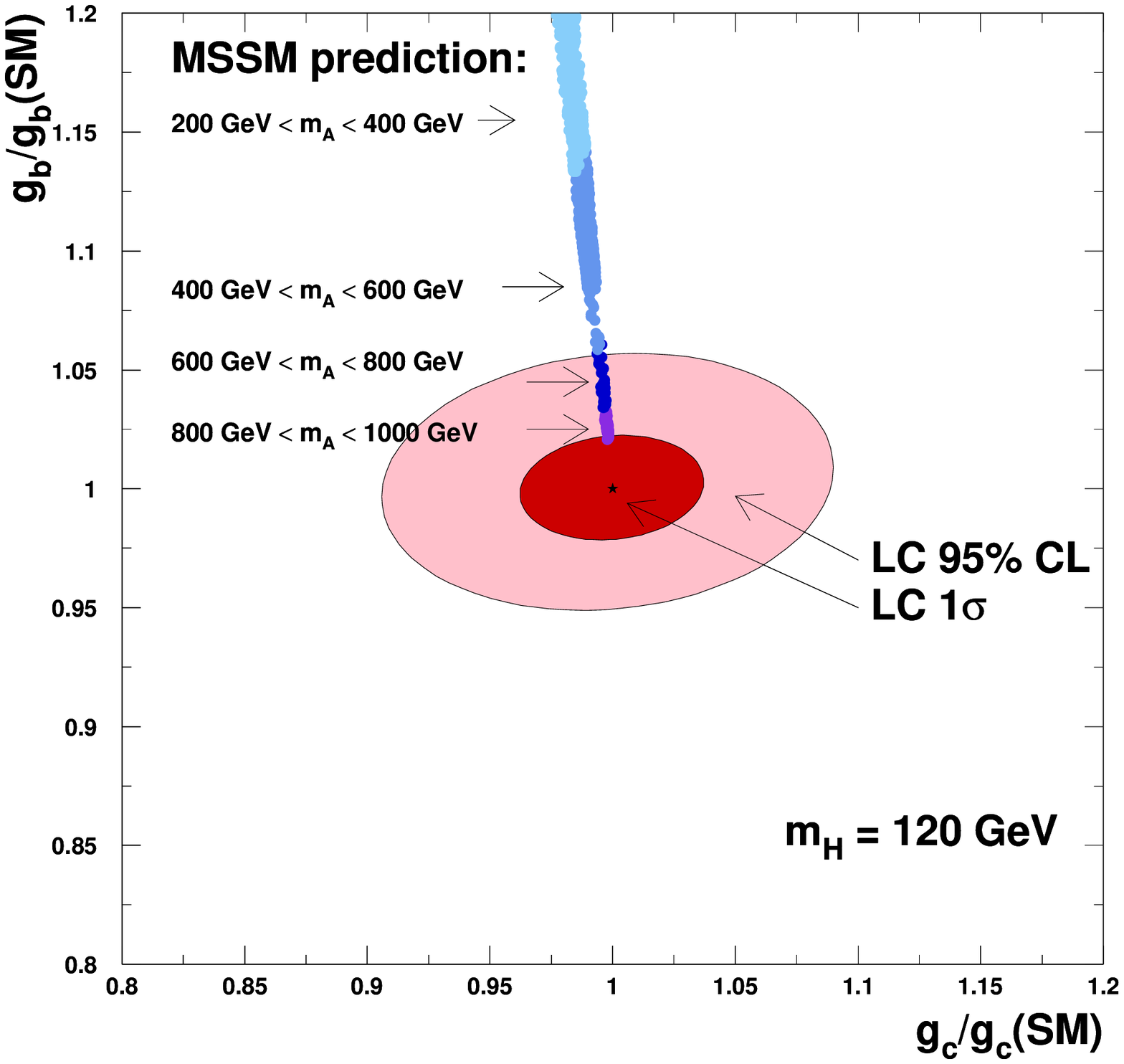,width=0.4\textwidth}
  {Expected sensitivity of the Higgs couplings $g_{\mathrm{Hbb}}$ vs.
    $g_{\mathrm{Hbb}}$ for a Higgs bosons mass 
    $m_{\mathrm{H}} = 120\,\GeV$ and $500\,\invfb$.
    \label{fig:hfitter}}
  Since the cross-section for this process is small 
  ($0.35\,\fb$ for $m_{\mathrm{H}}=120\,\GeV$ at 
  $\sqrt{s} = 500\,\GeV$), a very high luminosity is essential.
  A realistic study shows that $20\,\%$ precision of 
  $\lambda_{\mathrm{HHH}}$ seems possible for a light Higgs 
  boson \cite{HiggsTriH}.

  From the precision measurements of cross-sections and 
  branching ratios the couplings of the Higgs boson to 
  the different gauge bosons and fermions can be extracted 
  \cite{HiggsHfitter}. The expected precision ranges from 
  $1.2\,\%$ (for $g_{\mathrm{HZZ}}$) up to $3.7\,\%$ 
  (for $g_{\mathrm{Hc\bar{c}}}$) for a light Higgs boson 
  ($m_{\mathrm{H}} = 120\,\GeV$). As an example, Figure 
  \ref{fig:hfitter} shows the sensitivity of the couplings 
  $g_{\mathrm{Hbb}}$ and $g_{\mathrm{Hcc}}$ at a LC to the 
  mass of the pseudoscalar Higgs boson A in the MSSM.

  \section{Strong Electroweak Symmetry Breaking}

  Even if no light Higgs boson is found, the LC can contribute
  to the understanding of EWSB and the search for new physics.
  Unitarity requires that the interaction of gauge bosons becomes
  strong at high energies. Without knowing the details of the 
  new strong interaction, an effective Lagrangian can be used 
  to describe the interactions of electroweak gauge bosons 
  \cite{EWSBLagrangian}. Its coupling parameters $\alpha_1$, 
  $\alpha_2$ and $\alpha_3$ describe
  non-standard triple gauge boson couplings (TGC), and $\alpha_4$ 
  and $\alpha_5$ describe non-standard quartic gauge boson
  couplings (QGC), assuming $\mathrm{SU(2)}_c$ invariant and 
  linearly breaking operators, which conserve C and P.
  Each $\alpha_i$ is related to an energy 
  scale of new physics $\Lambda^*_i$ by
  \begin{equation}
    \label{eq:ewsb}
    \frac{\alpha_i}{16\pi^2} = \left(\frac{246\,\GeV}{\Lambda^*_i}\right)^2.
  \end{equation}
  The couplings are normalised in such a way that they are of $\mathcal{O}(1)$
  for new physics around $3\,\TeV$. 

  TGCs can be measured in the most general way in W-pair 
  production. A realistic  study of the semi-leptonic decay 
  channel $\PWp\PWm\to\Pq\Paq l\bar{\nu}_l$ at an energy of 
  $800\,\GeV$ shows that the expected errors for anomalous 
  TGCs \cite{Hagiwara:1987vm} $\Delta g^1_Z$, 
  $\Delta \kappa_{\gamma}$ and $\lambda_{\gamma}$ are in the 
  order of $10^{-4}$ \cite{TGCstudy}, which is up to two orders of 
  magnitude better than at LEP \cite{TGClep} or Tevatron. 
  This includes the use of polarised electrons ($\pm\, 80\,\%$) 
  and positrons ($\mp\, 60\,\%$). 

  \EPSFIGURE{sewsb_fig1.eps,width=0.535\textwidth}
  {Sensitivity for the strong EWSB parameters $\alpha_{1,2,3}$
    from a TGC measurement at $800\,\GeV$ and $1000\,\invfb$
    using polarised electrons ($\pm\, 80\,\%$) and positrons
    ($\mp\, 60\,\%$).
    \label{fig:sewsb}}

  The $\alpha$ parameters are a linear combination of $\Delta g^1_Z$, 
  $\Delta \kappa_{Z}$ and $\Delta \kappa_{\gamma}$. Unfortunately 
  this transformation is singular and only two out of three
  couplings can be measured independently in this process, 
  resulting in a blind direction in the parameter space. 
  Figure \ref{fig:sewsb} shows the expected errors using the 
  TGC measurement. The expected errors for $\alpha_2$ and 
  $\alpha_3$ can be interpreted by equation (\ref{eq:ewsb}) 
  as limits on the energy scale of new physics. Using electron 
  polarisation ($\pm\, 80\,\%$) yields $8.7\,\TeV$ for $\Lambda^*_2$ 
  and $11.6\,\TeV$ for $\Lambda^*_3$ at $800\,\GeV$
  at $68\,\%$ confidence level assuming $\Delta \alpha_1 = 0$. 
  These limits can be increased by a factor of 1.2 using also positron 
  polarisation ($\mp\, 60\,\%$).

  The parameter $\alpha_1$ can be independently obtained from
  electroweak precision observables at the \PZ-Pole. Figure 
  \ref{fig:sewsb} shows also the expected error on $\alpha_1$
  from a measurement at GigaZ. This option of a LC describes a high
  luminosity run at the \PZ-pole ($\sqrt{s}\approx 91\,\GeV$). A limit of 
  $5.9\,\TeV$ for $\Lambda^*_1$ at $68\,\%$ confidence level can be
  obtained.

  The QGC parameters $\alpha_4$ and $\alpha_5$ can be measured in 
  the processes $\epem\to\Pgne\Pagne\PWp\PWm$ and 
  $\epem\to\Pgne\Pagne\PZ\PZ$. A realistic study of the fully 
  hadronic decay channel was performed at a centre-of-mass energy 
  of $800\,\GeV$ \cite{QGCstudy}. The couplings are extracted 
  by fitting the total cross-section and differential distributions.
  The use of left-handed electrons ($-80\,\%$) and right-handed positrons 
  ($40\,\%$)
  improves the result significantly. The combination of both channels 
  removes ambiguities and increases the sensitivity. The expected errors 
  on $\alpha_4$ and $\alpha_5$ can again be interpreted as a limit on 
  the energy scale of new physics. Using equation (\ref{eq:ewsb}) 
  gives $2.3\,\TeV$ for $\Lambda^*_4$ and $3.1\,\TeV$ for $\Lambda^*_5$
  at $68\,\%$ confidence level. The limits improve asymptotically as
  1/$\sqrt{\mathcal{L}}$, where $\mathcal{L}$ is the integrated 
  luminosity.

  \section{Conclusion}

  An electron positron linear collider in the $90 - 1000\,\GeV$
  regime allows very significant insights into physics at the
  electroweak scale. The essential parts of the Higgs mechanism
  of electroweak symmetry breaking can be fully established 
  through precision measurements of Higgs boson properties.
  If no light Higgs boson is found, electroweak precision
  measurements of triple and quartic gauge boson couplings
  can extend the energy range for searches for new physics 
  far into the multi-TeV energy range.
  
\enlargethispage{5mm}

\end{document}